\documentclass[aps,preprint,a4paper,amsmath,showpacs,floatfix]{revtex4}

\usepackage{graphicx}
\usepackage{amssymb}
\usepackage{times}

\begin{document}

\title{A fundamental-measure density functional for mixtures of parallel
hard cylinders}

\author{Yuri Mart\'{\i}nez-Rat\'on,
Jos\'e A. Capit\'an and
Jos\'e A. Cuesta}

\affiliation{Grupo Interdisciplinar de Sistemas Complejos (GISC),
Departamento de Matem\'aticas, Escuela Polit\'ecnica Superior,
Universidad Carlos III de Madrid,
Avenida de la Universidad 30, E-28911 Legan\'es, Madrid, Spain.
}

\date{\today}

\begin{abstract}
We obtain a fundamental measure density functional for mixtures of 
parallel hard cylinders. To this purpose we first generalize to
multicomponent mixtures the fundamental measure functional proposed by
Tarazona and Rosenfeld for a one-component hard disk fluid, through
a method alternative to the cavity formalism of these authors. We
show the equivalence of both methods when applied to two-dimensional
fluids. The density functional so obtained reduces to the exact
density functional for one-dimensional mixtures of hard rods when
applied to one-dimensional profiles. In a second step we apply an
idea put forward some time ago by two of us, based again on a dimensional
reduction of the system, and derive a density functional for mixtures
of parallel hard cylinders. We explore some features of this functional
by determining the fluid-fluid demixing spinodals for a binary mixture
of cylinders with the same volume, and by calculating the direct correlation
functions.
\end{abstract}

% 61.20.Gy  Theory and models of liquid structure
% 61.30.Cz  Molecular and microscopic models and theories of liquid crystal structure 
% 64.75.+g  Solubility, segregation, and mixing; phase separation

\maketitle

\section{Introduction}

The fundamental measure density functional originally derived by Rosenfeld
for a fluid of hard spheres (HS)
\cite{rosenfeld:1986,rosenfeld:1988,rosenfeld:1989,rosenfeld:1990,rosenfeld:1992}
can be considered as the most sophisticated density functional (DF) that
has been successfully applied to the study of the highly confined HS fluid
\cite{gonzalez:1998,gonzalez:2006b} and to the HS freezing \cite{tarazona:2000}. 
The theoretical formalism initially developed by Rosenfeld to obtain 
the HS fundamental measure functional (FMF)
\cite{rosenfeld:1986,rosenfeld:1988,rosenfeld:1989} 
was later complemented with the concept of dimensional crossover to zero
dimension to obtain a DF that adequately describes the HS freezing
\cite{rosenfeld:1996,rosenfeld:1997}. By dimensional crossover it is understood
that when reducing the dimension of a $D$-dimensional system to that of a
$D'$-dimensional one (e.g.\ by confining), the $D$-dimensional DF crosses
over to the $D'$-dimensional one. This zero-dimensional (0D) crossover was
later employed by Tarazona and Rosenfeld to introduce a new cavity formalism
which showed how this unique property, together with the exact expressions 
for the zero- and one-dimensional (1D) HS functionals, are the only requirements 
needed to derive their final versions of the FMFs for the hard disk (HD) 
and the HS fluids \cite{tarazona:1997}. These versions which, as it is
standard in all FMFs, assume that the density profile dependence enters in
the functional only through a finite set of weighted densities, leave little
freedom for improvements without destroying the important dimensional 
crossover property. Recent efforts have been made to derive a HS functional
with an imposed equation of state (EOS) as its uniform limit
\cite{tarazona:2002,roth:2002}. This imposed EOS (for instance the
Carnahan-Starling EOS ) describes the fluid better than the scaled particle
result \cite{reis:1959} (the uniform limit of all the FMF derived from first
principles) in the description of the HS liquid. However, all these modifications
can be done at the expense of losing some, or all, of the dimensional crossovers
---a crucial property if one wants to study highly confined fluids. 

Thus, we can say that the fundamental measure theory (FMT) 
is close to its edge in the sense that it is questionable that 
any improvement can be achieved without rendering it intractable
\cite{cuesta:2002}. For instance, it has been argued that the inclusion
of an infinite set of weighted densities can remove some defects of the
HS FMF, because this is what happens ---with the addition of a few more weighted
densities--- for the FMF for parallel hard hexagons \cite{capitan:2007} (which
is constructed from the corresponding functional for parallel hard cubes   
\cite{cuesta:1996,cuesta:1997a,cuesta:1997b}). After all, a circle is a
polygon with infinitely many sides. Similar conclusions are reached when
FMFs for lattice models are constructed
\cite{lafuente:2002b,lafuente:2003a,lafuente:2003b}. In fact, the concept
of cavity can be generalized in those models so as to account for higher
orders of accuracy in the correlation functions \cite{lafuente:2004,lafuente:2005}.

The first extension of the FMF to general anisotropic particles was carried out
by Rosenfeld \cite{rosenfeld:1994b}. However this extension only works  
for isotropic fluids (with particle axes randomly oriented), because the Mayer
function is not recovered from the low-density expansion of the proposed FMF
when particles have a preferred alignment \cite{chamoux:1996}. In recent works
the definition of the one-particle weights necessary to calculate the weighted 
densities has been extended in such a way as to include an effective dependence
on the orientations of both interacting particles
\cite{schmidt:2001b,brader:2002,esztermann:2004,esztermann:2006}. This way 
the Mayer function can be exactly decomposed as a sum of convolutions between 
those extended weights. Different FMFs for freely rotating particles: for a
mixture of HS and hard needles \cite{schmidt:2001b,brader:2002}, and for a mixture
of hard needles and infinitely thin hard plates
\cite{esztermann:2004,esztermann:2006} have been proposed using these extended
weights, and the resulting FMFs have some of the desired dimensional crossovers.
However, the requirement of taking the 
breadth of the particles vanishing small seems to be indispensable to construct
such functionals. 

The parallel particle alignment (or the restricted orientation 
approximation) is a fundamental restriction that has to be taken
if we want to derive a FMF from first principles, without any 
approximation about the particle characteristic lengths. The FMF for 
parallel hard cubes \cite{cuesta:1996} and its later extension to hard 
parallelepipeds with orientation restricted to three perpendicular
axes (Zwanzig model) \cite{cuesta:1997a,cuesta:1997b} were the first examples
of FMFs for 
anisotropic particles derived form first principles using the original 
Rosenfeld's formalism and adding the dimensional cross-over constraint. 
The FMF for the Zwanzig model was applied to the calculation of 
phase diagrams of the one-component fluids made of hard rods and hard 
plates \cite{martinez-raton:2004},
and of phase diagrams of polydisperse rod-plate mixture 
\cite{martinez-raton:2002,martinez-raton:2003}. Also, it was applied to the study of 
interfacial phenomena in liquid crystals in three \cite{bier:2004}
and two \cite{martinez-raton:2007} dimensions.
Reference~\cite{harnau:2007} summarizes most of the works done on the study
of wall-liquid crystal fluid interfaces using the Zwanzig model. 

In this article we want to derive a FMF for another particle geometry 
with the parallel alignment restriction. In this case the fluid is composed 
of a mixture of parallel hard cylinders (PHCL) with different radii $R_i$ and
lengths $L_i$. To achieve this we will first of all extend the FMF obtained
by Tarazona and Rosenfeld \cite{tarazona:1997} for a one-component HD
fluid to a multicomponent mixture, using an alternative approach to
the cavity formalism used by these authors. We prove that the resulting
functional conforms all the dimensional crossovers and thus we show that 
both methods are completely equivalent. In a second step we derive a functional
for a mixture of parallel hard cylinders starting from the already obtained 
HD functional by applying a differential operator, as explained
in Ref.~\cite{cuesta:1997a}. This procedure guarantees the dimensional
crossover from 3D to 2D \cite{cuesta:1997a}.

The parallel alignment restriction is, of course, a hard constraint that 
prevents the use of the derived functional in the study of those phenomena
governed by changes in the orientational ordering of the constituent particles, 
as it often occurs in liquid-crystals. However, as the spatial correlations
are accurately treated, those phases with a high degree of orientation, such
as the smectic, columnar or crystalline phases at high pressures, should
be well described by the present functional. Also, the study of non-uniform
polydisperse liquid-crystalline phases, which are frequently present in
experiments on colloidal mixtures \cite{vanderkooij:2000a,vanderkooij:2000b},
is such a difficult task that the parallel alignment simplification seems 
to be the only way to take some steps forward in that direction.

The article is organized as follows. In Sec.~\ref{HD} we derive the FMF 
for mixtures of HD. In Sec.~\ref{cylinders} we use this result to construct
a FMF for mixtures of PHCL. Section~\ref{uniform1} discusses the uniform
mixture of PHCL and in Sec.~\ref{lac} we derive the direct correlation
functions for this mixture. After a section summarizing the results of the
paper, we include two appendices. In Appendix~\ref{2Dto1D} we proof that
the FMF for a mixture of HD has an exact 2D$\to$1D crossover. This is, of
course, inherited by the FMF for PHCL. In Appendix~\ref{c2} the
expressions for the geometric terms defining the direct correlation
functions are explicitly displayed.
   
\section{Fundamental measure density functional for a mixture of hard disks}
\label{HD}

In this section we will derive a density functional for mixtures of HD based
upon Tarazona and Rosenfeld's proposal for a one-component HD fluid obtained
by using the 0D cavity formalism of FMT \cite{tarazona:1997}. We will maintain
the functional structure of the excess part of the free energy density,
$\Phi^{\rm (2D)}({\bf r})$, and extend it to a multicomponent mixture by
calculating the kernel $K_{ij}(r)$ ($i$ and $j$ label disk species) which enters
the definition of the two-particle weighted density $N({\bf r})$ (see below). This
functional structure is
\begin{eqnarray}
\Phi^{\rm (2D)}({\bf r})=-n_0({\bf r})
\ln\left[1-n_2({\bf r})\right]+\frac{N({\bf r})}
{1-n_2({\bf r})},
\label{fi_2D}
\end{eqnarray}
and its extension to mixtures amounts to writing
\begin{eqnarray}
&&N({\bf r})=\sum_{i,j}\int d{\bf r}_1\int d{\bf r}_2\rho_i({\bf r}_1) 
\rho_j({\bf r}_2) \Omega_{ij}({\bf r}-{\bf r}_1,{\bf r}-{\bf r}_2),  
\label{N2}\\
&&\Omega_{ij}({\bf r}_1,{\bf r}_2)=\omega^{(0)}_i
({\bf r}_1)\omega^{(0)}_j({\bf r}_2)K_{ij}(r_{12}).
\label{N2a}
\end{eqnarray}
The one-particle weighted densities $n_0({\bf r})$ and $n_2({\bf r})$ 
[$n_2({\bf r})$ being the local packing fraction] are those of
Rosenfeld's for a FMF for HD mixtures
\cite{rosenfeld:1986,rosenfeld:1988,rosenfeld:1989}, i.e.   
\begin{eqnarray}
n_0({\bf r})&=&\sum_i\int d{\bf r}'\rho_i({\bf r}')\omega^{(0)}_i({\bf r}-
{\bf r}'),\label{n0}\\
n_2({\bf r})&=&\sum_i\int d{\bf r}'\rho_i({\bf r}')\omega^{(2)}_i({\bf r}-
{\bf r}'), 
\label{n2}
\end{eqnarray}
with $\rho_i({\bf r})$ the density profile of species $i$ and 
$\omega^{(\alpha)}_i({\bf r})$ the one-particle weights defined as 
\begin{eqnarray}
\omega^{(0)}_i({\bf r})=\frac{\delta(R_i-r)}{2\pi R_i}, \quad
\omega^{(2)}_i({\bf r})=\Theta(R_i-r),
\end{eqnarray}
$R_i$ ($i=1,2,\cdots,c$, with $c$ the number of components of the mixture) 
being the particle radii, and $\delta(x)$ and $\Theta(x)$ the 
Dirac delta and Heaviside step functions respectively. Equation~(\ref{N2})
is the natural extension to multicomponent mixtures of the two-particle
weighted density $N({\bf r})$ initially introduced for a one-component
fluid in Ref.~\cite{tarazona:1997}. The authors of this work found the
expression for the  kernel $K(r_{12})$ through the requirement that inserting
0D density profiles in the excess part of free energy 
\begin{equation}
\beta {\cal F}^{\rm (2D)}_{\rm{ex}}[\rho({\bf r})]=\int d{\bf r}\,
\Phi^{\rm (2D)}({\bf r})
\label{eq:Fex2D}
\end{equation}
should recover the interaction part of the free energy of a 0D cavity
$\Phi^{\rm (0D)}={\cal N}+(1-{\cal N})\ln(1-{\cal N})$, with ${\cal N}<1$ 
the mean occupation of the cavity. As usual, the 0D cavity is understood
as a cavity of arbitrary geometry which can accommodate one particle at
the most.

\begin{figure}
\includegraphics[width=2.in]{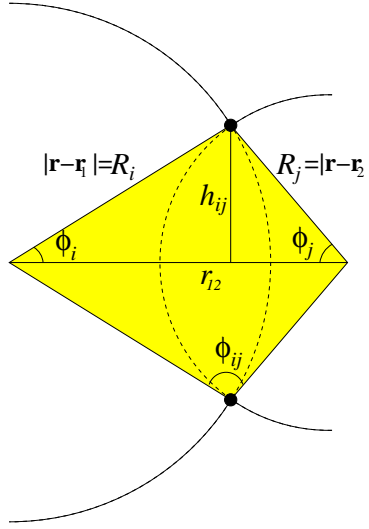}
\caption{Sketch of the triangular  geometry defined by 
the lengths $R_i$, $R_j$ and $r_{12}$.} 
\label{sketch1}
\end{figure}

To determine $K_{ij}(r)$ we will follow another procedure: we will
impose that the low density limit of the second functional 
derivative of (\ref{eq:Fex2D}) with respect to the density profiles 
$\rho_i({\bf r}_1)$ and $\rho_j({\bf r}_2)$ coincides with the overlap
function of two HD of radii $R_i$ and $R_j$, which turns out to be
the exact low density limit of minus the direct correlation function, i.e.
\begin{eqnarray}
\Theta(R_{ij}^{(+)}-r_{12})=
\int d{\bf r'}\left\langle\omega^{(0)}_i({\bf r}')\omega_j^{(2)}
({\bf r}_{12}-{\bf r}')\right\rangle+
2K_{ij}(r_{12})\int d{\bf r}'\omega^{(0)}_i({\bf r}')\omega^{(0)}_j
({\bf r}_{12}-{\bf r}'),
\label{lineal}
\end{eqnarray}
where $\langle f_{ij}\rangle=f_{ij}+f_{ji}$ has been introduced to
denote symmetrization of $f_{ij}$ with respect to its indices, and
$R_{ij}^{(+)}=R_i+R_j$. The calculation of the integrals involved in
Eq.~(\ref{lineal}) leads to
\begin{eqnarray}
K_{ij}(r_{12})&=&\pi r_{12}\langle \sin^{-1}t_{ij}(r_{12})\rangle
R_i\sqrt{1-t_{ij}(r_{12})^2}\,\Theta\left(r_{12}-R_{ij}^{(-)}\right)
\Theta\left(R_{ij}^{(+)}-r_{12}\right), \label{kernel}\\ 
t_{ij}(r_{12})&=&\frac{r_{12}^2+R_i^2-R_j^2}{2r_{12}R_i},
\label{kernel2}
\end{eqnarray}
where $R_{ij}^{(-)}=\left|R_i- R_j\right|$. The kernel (\ref{kernel}) 
is symmetric with respect to the exchange of indices $i$ and $j$  
due to the equality $h_{ij}\equiv R_i\sqrt{1-t_{ij}(r_{12})^2}=
R_j\sqrt{1-t_{ji}(r_{12})^2}$, which is easily visualized in
Fig.~\ref{sketch1}: this figure shows a sketch of a typical configuration
of two HD with different radii for which the $K_{ij}(r_{12})$ is different
from zero; in it the height $h_{ij}$ of the triangle formed by the lengths
$r_{12}$, $R_i$ and $R_j$ can be calculated either as
$R_i\sin\phi_i$ or as $R_j\sin\phi_j$, thus proving the symmetry
$K_{ij}(r_{12})=K_{ji}(r_{12})$. Also, from Eq.~(\ref{kernel}) and the 
triangular geometry the kernel can be rewritten as 
$K_{ij}(r_{12})=\pi A_{ij}(r_{12})\phi_{ij}(r_{12})$, 
with $A_{ij}=r_{12}h_{ij}$ the sum of the areas of both 
triangles (the shaded region of Fig.~\ref{sketch1}), 
and $\phi_{ij}=\langle \sin^{-1} t_{ij}\rangle$,
the angle formed by the sides $R_i$ and $R_j$.     

\begin{figure}
\includegraphics[width=3.in]{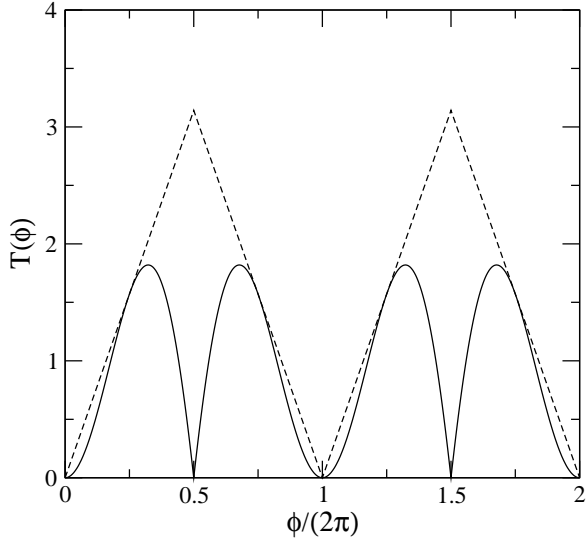}
\caption{The function $T(\phi)$ in the interval 
$[0,4\pi]$ (solid line). Also is shown with dashed line 
the triangular wave function.}  
\label{triangular}
\end{figure}

For one component, $K_{ij}(r)$ recovers the expression  
\begin{equation}
K(r)=4\pi R^2 \left(\frac{r}{2R}\right)
\sin^{-1}\left(\frac{r}{2R}\right)\sqrt{1-\left(\frac{r}{2R}\right)^2}
\end{equation}
reported in Ref.~\cite{tarazona:1997}.
It is interesting to note that, because of the presence of  
weights $\omega^{(0)}_i({\bf r})$ (proportional to Dirac delta 
functions) in the definition of $N({\bf r})$, its expression  
can be greatly simplified. After insertion of $K_{ij}(r)$ into 
Eq.~(\ref{N2}) and integration over the radial variables $r_1$ and $r_2$
we find  
\begin{eqnarray}
N({\bf r})=\frac{1}{4\pi}\sum_{i,j}R_iR_j
\int_0^{2\pi} d\phi_1\int_0^{2\pi} d\phi_2\rho_i({\bf r}+R_i{\bf u}_1)
\rho_j({\bf r}+R_j{\bf u}_2)T(\phi_{12}),
\label{new_N2}
\end{eqnarray}
where $\phi_{12}=\phi_1-\phi_2$, ${\bf u}_i=(\cos\phi_i,\sin\phi_i)$, 
and the function $T(\phi)$ is defined as 
\begin{eqnarray}
T(\phi)=|\phi-2\pi n||\sin\phi|, 
\label{wave}
\end{eqnarray}
with $n$ 
the integral part of the fraction $(\phi+\pi)/(2\pi)$. 
The first factor on the right hand side of (\ref{wave}) is the
$2\pi$-periodic function shown in Fig.~\ref{triangular} with a dashed line.
The function $T(\phi)$ is plotted in the same figure. From the new form
of $N({\bf r})$ given by Eq.~(\ref{new_N2}) we can conclude that, despite
the presence of a two-particle weight in its definition,
the numerical cost required to evaluate it is the same as that required
to calculate the local packing fraction $n_2({\bf r})$, because
both quantities are defined through a double integral.

Taking into account the result 
$\int d\phi_1\int d\phi_2T(\phi_{12})=4\pi^2$ we find that the uniform 
limit of $\Phi^{\rm (2D)}$ [cf.~Eq.~(\ref{fi_2D})] 
coincides with the scaled particle theory (SPT) result for a 
mixture of HD \cite{reis:1959,rosenfeld:1986,rosenfeld:1988,rosenfeld:1989} 
\begin{eqnarray}
\Phi^{\rm (2D)}
=-\xi_0\ln(1-\xi_2)+\frac{1}{4\pi}\frac{\xi_1^2}{1-\xi_2},
\label{SPT}
\end{eqnarray}
where $\xi_0=\sum_i\rho_i$, $\xi_1=\sum_i\rho_i (2\pi R_i)$, and 
$\xi_2=\sum_i\rho_i \left(\pi R_i^2\right)$. 

Because the derivation of $K_{ij}(r)$ has not followed the requirement of
exact reduction to 0D cavities, as in Ref.~\cite{tarazona:1997}, the question
arises as to whether this nice property holds for this new functional.
In Appendix~\ref{2Dto1D} we proof a stronger property: the functional
(\ref{fi_2D}) fulfills an exact 2D$\to$1D crossover; in other words, by
inserting the profile $\rho_i({\bf r})=\rho_i(x)\delta(y)$ into
(\ref{fi_2D}) and (\ref{eq:Fex2D}) we recover the exact functional for
1D hard rod mixtures \cite{vanderlick:1989}, cf.~Eqs.~(\ref{eq:phi1D})
and (\ref{eq:1Dfunctional}). With this we have also proven that
the present method and the cavity formalism of Ref.~\cite{tarazona:1997}
are two equivalent methods to obtain a FMF for mixtures of HD.  

\section{Density functional for a mixture of parallel hard cylinders}
\label{cylinders}

In this Section we will construct a FMF for a mixture of PHCL
starting from two different density functionals for HD mixtures: the
first one is calculated through Eq.~(\ref{fi_2D}), while
the second one is Rosenfeld's proposal \cite{rosenfeld:1989}    
\begin{eqnarray}
\Phi^{\rm (2D)}_R=-n_0\ln(1-n_2)+\frac{1}{4\pi}\frac{v^2-{\bf v}^2}{1-n_2},
\label{Yasha}
\end{eqnarray}
where $v=\sum_i\rho_i\ast w_i({\bf r})$ and
${\bf v}=\sum_i\rho_i\ast {\bf w}_i({\bf r})$, with the new weights 
\begin{eqnarray}
w_i({\bf r})=2\pi R_i\omega_i^{(0)}({\bf r}), 
\qquad {\bf w}_i({\bf r})=w_i \frac{{\bf r}}{r}.
\end{eqnarray}  
Note that the weighted density $v({\bf r})$ is not 
the absolute value of ${\bf v}({\bf r})$. 
Rosenfeld obtained this expression by approximating 
the Mayer function of two HD by the sum of convolutions 
between single-particle weights
$f_{ij}(r)=\langle\omega^{(0)}_i\ast\omega^{(2)}_j\rangle(r)+(2\pi)^{-1}
[w_i\ast w_j(r)-{\bf w}_i\ast{\bf w}_j(r)]$
and requiring also that the scaled particle result (\ref{SPT}) was recovered 
in the uniform limit \cite{rosenfeld:1989}. The expression (\ref{Yasha}),
however, does not conform with any dimensional crossover to lower dimensions. 

The FMF for PHCL is obtained by resorting to the dimensional crossover
3D$\to$2D which any functional $\beta{\cal F}^{\rm (3D)}_{\rm{ex}}[\{\rho_i\}]$
fulfill. In Ref.~\cite{cuesta:1997a} it was argued that having a FMF for a
mixture of two-dimensional particles ${\cal F}_{\rm{ex}}^{\rm (2D)}
[\{\rho_i\}]$ one can construct an explicit expression for a FMF for a
mixture of parallel anisotropic three-dimensional (3D) bodies whose
constant section perpendicular to their main axes is that of the 2D
particles. In the same reference it is shown that the resulting functional
fulfills by construction the 3D$\to$2D dimensional crossover when the original
fluid is confined such that the centers of mass of the particles are confined
on a plane perpendicular to their axes. We will apply this method to obtain
a FMF for a mixture of parallel cylinders as follows. First of all we need
to redefine the functional in such a way as to include the $z$-coordinate
dependence of the density profiles and correspondingly of the weighted
densities. The new weights are obtained multiplying  
the old ones by the factors $\Theta(L_i/2-|z|)$, i.e. 
\begin{eqnarray}
&&\omega^{(1)}_i({\bf r})=\omega^{(0)}_i({\bf r}^{\perp})
\Theta(L_i/2-|z|), \quad 
\omega^{(3)}_i({\bf r})=\omega^{(2)}_i({\bf r}^{\perp})\Theta(L_i/2-|z|), 
\label{w1w3}\\ 
&&\Omega^{(2)}_{ij}({\bf r}_1,{\bf r}_2)=\Omega_{ij}
({\bf r}_1^{\perp},{\bf r}_2^{\perp})\Theta(L_i/2-|z_1|)
\Theta(L_j/2-|z_2|).\label{Omega}
\end{eqnarray}
The resulting free energy density, which we denote $\tilde{\Phi}^{\rm (2D)}$,
is the same as that given by Eq.~(\ref{fi_2D}), but with
the substitutions $n_0({\bf r})\to n_1({\bf r})$, 
$n_2({\bf r})\to n_3({\bf r})$ and $N({\bf r})\to N_2({\bf r})$. The 
new weighted densities are obtained through the same expressions given for
the 2D case but using the new weights (\ref{w1w3}), (\ref{Omega}). Note
that the vector position ${\bf r}$ is now defined as 
${\bf r}=({\bf r}^{\perp},z)$. The dimensional crossover 3D$\to$2D holds
if the 3D excess free-energy density is obtained by the following formula 
\cite{cuesta:1997a,cuesta:1997b}
\begin{eqnarray}
\Phi^{\rm (3D)}({\bf r})=\sum_i\frac{\partial}{\partial L_i}
\tilde{\Phi}^{\rm (2D)}({\bf r}).
\label{almendruco}
\end{eqnarray}
In our case this leads to
\begin{eqnarray}
\Phi^{\rm (3D)}=-n_0\ln(1-n_3)+\frac{n_1n_2+N_1}{1-n_3}+\frac{n_2N_2}
{(1-n_3)^2},
\label{nuestro}
\end{eqnarray}
where the one-particle weighted densities 
$n_{\alpha}({\bf r})$ are  calculated as usual as 
$n_{\alpha}({\bf r})=\sum_i\rho_i\ast\omega^{(\alpha)}_i({\bf r})$
[$\ast$ stands for the convolution $f\ast g({\bf r})\equiv 
\int d{\bf r}'f({\bf r}')g({\bf r}-{\bf r}')$]. 
The one-particle weights are defined jointly by 
Eq. (\ref{w1w3}) and  
\begin{eqnarray}
\omega^{(0)}_i({\bf r})=\frac{1}{2}\omega^{(0)}_i({\bf r}_{\perp})
\delta(L_i/2-|z|),\quad \omega_i^{(2)}({\bf r})=\frac{1}{2}
\omega^{(2)}_i({\bf r}_{\perp})\delta(L_i/2-|z|).
\label{eq:w0w2}
\end{eqnarray}
The two-particle weighted densities are calculated as 
\begin{eqnarray}
N_{\alpha}({\bf r})=\sum_{ij}\int d{\bf r}_1\int d{\bf r}_2 \rho_i({\bf r}_1)\rho_j({\bf r}_2)
\Omega^{(\alpha)}_{ij}({\bf r}-{\bf r}_1,{\bf r}-{\bf r}_2), 
\end{eqnarray}
with $\Omega_{ij}^{(2)}({\bf r}_1,{\bf r}_2)$ given by Eq. (\ref{Omega}) while 
$\Omega_{ij}^{(1)}({\bf r}_1,{\bf r}_2)$ is  
\begin{eqnarray}
\Omega^{(1)}_{ij}({\bf r}_1,{\bf r}_2)=\Omega_{ij}
({\bf r}_1^{\perp},{\bf r}_2^{\perp})
\left\langle \frac{1}{2}\Theta(L_i/2-|z_1|)\delta(L_j/2-|z_2|)\right\rangle.
\label{Omega1}
\end{eqnarray}

As mentioned above, the free-energy functional ${\cal F}^{\rm (3D)}$ has a
correct dimensional reduction to ${\cal F}^{\rm (2D)}$, the free energy for a 
HD mixture, when the density profiles are taken as $\rho_i({\bf r})=
\rho^{\rm (2D)}_i\left({\bf r}^{\perp}\right)\delta(z)$ 
$[{\bf r}^{\perp}=(x,y)$], i.e. projecting 
the cylinders on the plane perpendicular to their axes. It is easy 
to show that the dimensional cross-over 3D$\to$2D, where the 
projection is now in a plane parallel to the cylinder 
axes, also holds. To show this we take the density profiles 
as $\rho_i({\bf r})=\rho_i({\bf r}^{\parallel})\delta(x)$ 
[${\bf r}^{\parallel}=(y,z)$], with 
$\rho_i({\bf r}^{\parallel})=\rho_i^{\rm (2D)}(y,z)$.
Inserting these $\rho_i({\bf r})$ into $\tilde\Phi^{\rm (2D)}({\bf r})$
and using the already shown dimensional cross-over 2D$\to$1D 
of a FMF for a mixture of HD (see Appendix~\ref{2Dto1D}), we obtain,
from Eq.~(\ref{almendruco}),
\begin{eqnarray}
\Phi^{\rm (3D)}({\bf r})\to \Phi^{\rm (2D)}_{\rm{PHR}}({\bf r})=
-n_0({\bf r})\ln\left[1-n_2({\bf r})\right]+\frac{n_{1x}({\bf r})n_{1y}
({\bf r})}
{1-n_2({\bf r})},  
\end{eqnarray}
the free-energy density of a mixture of parallel hard rectangles 
(the section of the cylinders along their axes) \cite{cuesta:1997a,cuesta:1997b}. 
The weighted densities for such particles are now defined as  
$n_0({\bf r})=\sum_i\rho_i^{\rm (2D)}\ast \left[\delta_{\sigma_i}(y)
\delta_{L_i}(z)\right]$, 
$n_2({\bf r})=\sum_i\rho_i^{\rm (2D)}\ast\left[\theta_{\sigma_i}(y)
\theta_{L_i}(z)\right]$,  
$n_{1x}({\bf r})=\sum_i\rho_i^{\rm (2D)}
\ast\left[\delta_{\sigma_i}(y)\theta_{L_i}(z)\right]$, and 
$n_{1y}({\bf r})=\sum_i\rho_i^{\rm (2D)}\ast\left[\theta_{\sigma_i}(y)
\theta_{L_i}(z)\right]$, where 
the shorthand notations $\delta_{u_i}(s)=\frac{1}{2}\delta(u_i/2-|s|)$ 
and $\theta_{u_i}(s)=\Theta(u_i/2-|s|)$, with $s=y,z$ and $\sigma_i=2R_i$,
have been used.

\begin{figure}
\includegraphics[width=155mm]{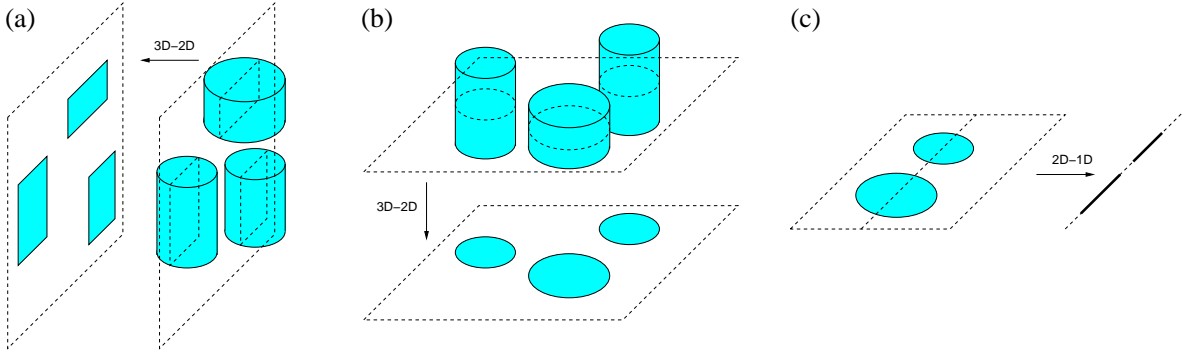}
\caption{Sketch of all dimensional crossovers fulfilled by the fundamental-measure
functional for a mixture of parallel hard cylinders: (a) from cylinders (3D) to
rectangles (2D), (b) from cylinders (3D) to disks (2D) and (c) from disks (2D)
to rods (1D).}
\label{cross-over}
\end{figure}

Thus the FMF for PHCL that we have just obtained conforms
with all dimensional crossovers to lower dimensions, which we have
sketched if Fig.~\ref{cross-over}. Nevertheless,
this functional is not perfect because it shares with that of HD the
defect caused by the existence of ``lost cases'' \cite{tarazona:1997,cuesta:2002}.
There are three-point 0D cavities such that particles sited at those
points have pairwise overlap but no triple overlap.
For those cavities the FMF of a HD mixture does not reduce 
adequately to $\Phi^{\rm (0D)}$ (the lost cases for the one-component 
fluid were already pointed out in Ref.~\cite{tarazona:1997}). As a
consequence of that the FMF for a mixture of PHCL suffers from the
same illness.

The excess free-energy density of PHCL obtained from its HD counterpart 
using (\ref{almendruco}) and Rosenfeld' approximation $\Phi^{\rm (2D)}_R$
[cf.~Eq.~(\ref{Yasha})] results in 
\begin{eqnarray}
\Phi^{\rm (3D)}_R=-n_0\ln(1-n_3)+
\frac{n_1n_2+v_1v_2-{\bf v}_1{\bf v}_2}{1-n_3}
+\frac{1}{4\pi}\frac{n_2\left(v_2^2-{\bf v}_2^2\right)}{(1-n_3)^2},
\label{Yasha1}
\end{eqnarray}
where $v_{\alpha}({\bf r})=\sum_i\rho_i\ast w_i^{(\alpha)}({\bf r})$, 
${\bf v}_{\alpha}({\bf r})=\sum_i\rho_i\ast{\bf w}_i^{(\alpha)}({\bf r})$, and 
\begin{align}
w_i^{(1)}({\bf r})&=R_i\omega^{(0)}_i({\bf r}), \quad  &
{\bf w}_i^{(1)}({\bf r}) &=w_i^{(1)}({\bf r})\frac{{\bf r}^{\perp}}{R_i},\\
w_i^{(2)}({\bf r})&=2\pi R_i\omega^{(1)}_i({\bf r}), \quad &
{\bf w}_i^{(2)}({\bf r}) &=w_i^{(2)}({\bf r})\frac{{\bf r}^{\perp}}{R_i},
\end{align}
where the $\omega_i^{(\alpha)}({\bf r})$ are those defined by
Eqs.~(\ref{w1w3}) and (\ref{eq:w0w2}).

\section{Uniform mixtures}
\label{uniform1}

In this section we give the explicit expression for the uniform limit
of the FMF for a mixture of PHCL, which coincides with the SPT result.  

It is easy to show that $\int d{\bf r}_1\int d{\bf r}_2 \Omega_{ij}
({\bf r}_1,{\bf r}_2)=\pi R_iR_j$. Taking into account this result,
the uniform limit $\rho_i({\bf r})=\rho_i$ of both free-energy 
densities, $\Phi^{\rm (3D)}$ from Eq.~(\ref{nuestro}) and $\Phi^{\rm (3D)}_R$ 
from Eq.~(\ref{Yasha1}), yield the result
\begin{eqnarray}
\Phi^{\rm (3D)}_u=-\xi_0\ln(1-\xi_3)+
\frac{\boldsymbol{\xi}_1\cdot\boldsymbol{\xi}_2}{1-\xi_3}+
\frac{1}{8\pi}\frac{\xi_2^{\parallel}\left(\xi_2^{\perp}\right)^2}{
(1-\xi_3)^2},
\label{uniform}
\end{eqnarray}
where we have defined the vectors $\boldsymbol{\xi}_i\equiv
\left(\xi_i^{\perp},\xi_i^{\parallel}\right)$ ($i=1,2$), with components  
\begin{align}
\xi_1^{\perp}&=\sum_i\rho_i R_i, \qquad & \xi_1^{\parallel} &=
\sum_i\rho_i \frac{L_i}{2}, \label{ala1}\\
\xi_2^{\perp}&=\sum_i\rho_i 2\pi R_iL_i,\qquad &
\xi_2^{\parallel} &=\sum_i\rho_i2\pi R_i^2,
\label{ala2}
\end{align} 
while $\xi_0=\sum_i\rho_i$ and $\xi_3=\sum_i\rho_i \pi R_i^2L_i$ are 
the total density and total packing fraction of the mixture, respectively.
{}From Eqs. (\ref{uniform})--(\ref{ala2}) we can see that the 
excess part of free-energy density is a function of certain weighted 
densities $\xi_i^{(\alpha)}$, which can be calculated as the sum of  
products between the particle densities 
$\rho_i$ and their fundamental measures: $\{R_i,L_i/2\}$, the 
principal radii in the directions perpendicular and parallel to the
cylinder axes, $\{2\pi R_iL_i,2\pi R_i^2\}$, the areas of the surfaces
oriented along the perpendicular and parallel 
directions, and $\pi R_i^2L_i$ the particle volume.  

Within the SPT formalism, the excess part of the free-energy density
of any mixture of convex particles should fulfill the following differential 
equation \cite{rosenfeld:1989,cuesta:1997b}
\begin{eqnarray}
-\Phi+\sum_i\xi_i\frac{\partial \Phi}{\partial \xi_i}+\xi_0=
\frac{\partial \Phi}{\partial \xi_3}.
\label{diff}
\end{eqnarray} 
This equation holds for (\ref{uniform}), thus showing that of our
functional gives the SPT result. Finally, the equation of state within SPT 
can be calculated as 
$\beta P=\displaystyle{\frac{\partial\Phi_u}{\partial\xi_3}}$, resulting in
\begin{eqnarray}
\beta P=\frac{\xi_0}{1-\xi_3}+\frac{\boldsymbol{\xi}_1\cdot\boldsymbol{\xi}_2}
{(1-\xi_3)^2}+\frac{1}{4\pi}
\frac{\xi_2^{\parallel}\left(\xi_2^{\perp}\right)^2}{(1-\xi_3)^3}.
\label{EOS}
\end{eqnarray}
This equation of state can be used to study the possible demixing 
scenarios that a mixture of PHCL has.

In order to show the existence of 
demixing in a binary mixture of PHCL we first specialize the 
mixture to the case in which both particle volumes are unity, 
i.e.\ $v_1=v_2=1$. This assumption allows us to calculate the particle
lengths and diameters as $L_i=c \kappa_i^{2/3}$ and 
$\sigma_i=c\kappa_i^{-1/3}$, where $\kappa_i=L_i/\sigma_i$ are the
cylinders aspect ratios and $c=(4/\pi)^{1/3}$. It is easy to show 
that the expression (\ref{uniform}) for this particular mixture gives us 
the following expression for the free-energy per particle 
$\varphi=\left(\Phi_{\rm{id}}+\Phi^{\rm (3D)}_u\right)/\rho$ (with 
$\Phi_{\rm{id}}=\sum_i \rho_i\left[\ln(v_i\rho_i)-1\right]$)
\begin{eqnarray}
\varphi=\ln y-1+x\ln x+(1-x)\ln(1-x)+y S(x;r)+y^2T(x;r),
\label{varphi}
\end{eqnarray}
while the expression for the fluid pressure is 
\begin{eqnarray}
\beta Pv_i=y+y^2S(x;r)+2y^3T(x;r),
\end{eqnarray}
where $y\equiv \eta/(1-\eta)$, $x\equiv x_2$ is the molar fraction of 
species 2 and $r\equiv \kappa_2/\kappa_1$ is the ratio between the
particles aspect ratios. Note that for this particular mixture we have 
$\eta=\rho$. Also,
\begin{eqnarray}
S(x;r)&=&3+\left(r^{1/3}-1\right)^2\left(1+4r^{-1/3}+r^{-2/3}\right)
x(1-x),\\
T(x;r)&=&1+\left(r^{1/3}-1\right)^2\left[r^{-1/3}\left(2+r^{-1/3}\right)+
\left(1-r^{-2/3}\right)x\right]x(1-x).
\end{eqnarray}
Note that while the function $S(x;r)=S(1-x;r)$ 
is symmetric with respect to the value 
$x=1/2$, $T(x;r)$ is not. Thus the spinodal instability curve 
with respect to the phase separation is not symmetric with respect to 
$x=1/2$. Besides we have the obvious symmetry $S(x;r)=S(1-x;r^{-1})$ and 
$T(x;r)=T(1-x;r^{-1})$.

\begin{figure}
\includegraphics[width=3.5in]{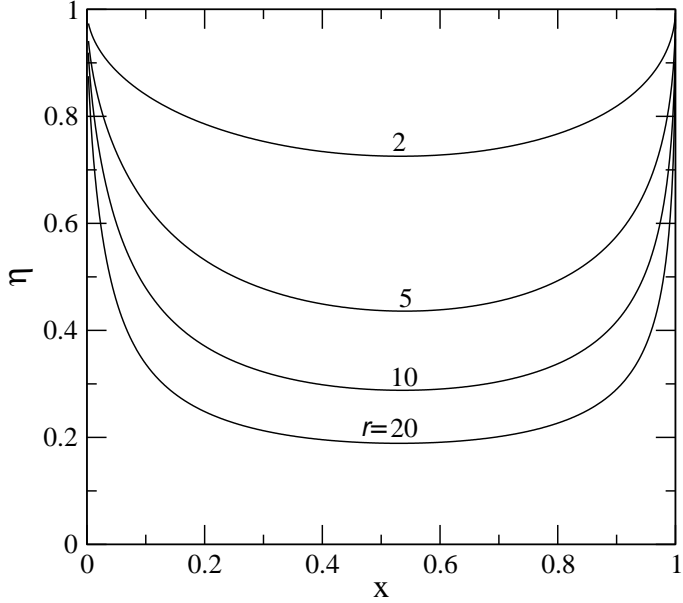}
\caption{Demixing spinodals for the phase separation between two 
nematic phases of different composition calculated for values 
of the asymmetry parameter $r$ as labeled in the figure 
.} 
\label{demix}
\end{figure}

The lost of mixture stability with respect to phase segregation 
can be calculated as usual as 
\begin{eqnarray}
\text{det}\left(\rho_i^{-1}\delta_{ij}+
\frac{\partial^2\Phi^{\rm (3D)}_u}{\partial\rho_i\partial\rho_j}\right)=0,
\end{eqnarray} 
which is equivalent to the following condition, expressed in the 
variables $y$ and $x$,
\begin{eqnarray}
\frac{\partial }{\partial y}\left[y^2\frac{\partial\varphi}{\partial y}\right]
\frac{\partial^2\varphi}{\partial x^2}-\left(y\frac{\partial^2
\varphi}{\partial y\partial x}\right)^2=0.
\label{cond}
\end{eqnarray}
Inserting (\ref{varphi}) into (\ref{cond}) we
calculate the demixing spinodals for different values of the 
asymmetry parameter $r$ in the plane $x-\eta$. 
Figure~\ref{demix} shows these 
demixing spinodals for values $r=20$, 10, 5 and $2$.

Of course this analysis does not prove that a thermodynamically stable
fluid-fluid demixing occurs, as inhomogeneous phases are not being
accounted for.

\section{Direct correlation function}
\label{lac}

The second functional derivative of $\beta {\cal F}^{\rm (3D)}_{\rm{ex}}=
\int d{\bf r}\,\Phi^{\rm (3D)}({\bf r})$ with respect to the density profiles 
$\rho_i({\bf r})$ and $\rho_j({\bf r})$ evaluated at the uniform 
densities $\rho_i$ gives us, after a long and tedious calculation,
the following expression for the direct correlation function
\begin{eqnarray}
c_{ij}({\bf r}_{12})=\left[\chi_0+\boldsymbol{\chi}_1\cdot\Delta{\bf R}_{ij}
({\bf r}_{12})+\boldsymbol{\chi}_2\cdot\Delta{\bf S}_{ij}({\bf r}_{12})
+\chi_3\Delta V_{ij}({\bf r}_{12})\right]f_{ij}(r_{12}),
\label{corr}
\end{eqnarray}
where $\displaystyle{\chi_i=\frac{\partial (\beta P)}{\partial \xi_i}}$ and 
\begin{eqnarray}
f_{ij}(r_{12})&=&-\Theta\left(R_{ij}^{(+)}-r_{12}\right)\Theta\left(
L_{ij}^{(+)}/2-|z_{12}|\right),\label{mayer}\\
\Delta {\bf R}_{ij}({\bf r}_{12})&=&
\left[\Delta L_{ij}^{\perp}(r_{12}^{\perp})/(2\pi),
\Delta L_{ij}^{\parallel}(z_{12})/2\right],\label{radii}\\
\Delta {\bf S}_{ij}({\bf r}_{12})&=&\left[
\Delta S_{ij}^{\perp}({\bf r}_{12}),\Delta S_{ij}^{\parallel}(r_{12}^{\perp})
\right],\label{surface}\\
\Delta V_{ij}({\bf r}_{12})&=&\frac{1}{2}\Delta S_{ij}^{\parallel}(r_{12}^{\perp})
\Delta L_{ij}^{\parallel}(z_{12})\label{volume},
\label{vol}
\end{eqnarray}
with $L_{ij}^{(\pm)}\equiv|L_i\pm L_j|$,
are the Mayer function (\ref{mayer}) and the geometrical measures of the body 
defined by the overlap 
between two cylinders $i$ and $j$ whose centers of mass are 
separated by the vector ${\bf r}_{12}$. These 
measures are characteristic radii along the perpendicular and parallel  
directions (\ref{radii}), the oriented surfaces (\ref{surface}) 
and the total overlap volume (\ref{volume}). The 
radii in turn are defined through the total length $L_{ij}^{(\perp)}$ of 
the arches and the height   
$L_{ij}^{(\parallel)}$ of the overlap body. The expressions for 
these quantities as well as for
$\Delta S_{ij}^{(\alpha)}$ ($\alpha=\perp,\parallel$) are given 
in Appendix~\ref{c2}.

The form of the direct correlation function (\ref{corr}) as a function of 
the geometric measures of the overlap body is exactly the same 
as that obtained from the Percus-Yevick approximation for a HS mixture, as it
was first shown by Rosenfeld \cite{rosenfeld:1988,rosenfeld:1989}.
The same formal expression is also obtained for a mixture of parallel hard
cubes \cite{cuesta:1997a,cuesta:1997b}. 

\section{Conclusions}

We have derived a FMF for a mixture of HD, and further used it to construct
another one for a mixture of PHCL. The resulting functional fulfills all
dimensional crossovers, a feature that makes the obtained functional very
useful in the study of fluid mixtures of perfectly aligned hard rods
confined by external potentials. These external potentials may 
have planar or cylindrical geometry. Some interfacial phase transitions,
such as wetting, layering and capillary ordering, can be studied as well
using this functional.

Of course, the parallel alignment constraint limits the use of the PHCL
functional to the study of highly oriented phases, such as nematic,
smectic or crystal phases at very high pressures. A particularly interesting
application of this functional is the determination of the phase behavior
of polydisperse hard rod mixtures. The inclusion of smectic and columnar
phases in the study makes the constraint of perfect particle alignment
indispensable to achieve the numerical minimization of the functional. 
Some experimental works \cite{vanderkooij:2000a,vanderkooij:2000b} as well
as simulations \cite{polson:1997} predict that polydispersity enhances the
columnar phase stability with respect to the smectic phase. It will be
interesting to check these conclusions with the reported functional.

\section*{ACKNOWLEDGMENTS}

Y.\ Mart\'{\i}nez-Rat\'on is supported by a Ram\'on y Cajal research contract. 
J.\ A.\ Capit\'an acknowledges
financial support through a contract from Consejer\'{\i}a de
Educaci\'on of Comunidad de Madrid and Fondo Social Europeo.
This work is part of research projects MOSAICO of the Ministerio de
Educaci\'on y Ciencia (Spain), and MOSSNOHO of Comunidad 
Aut\'onoma de Madrid (Spain).

\appendix

\section{2D$\to$1D limit of the FMF for a mixture of hard disks}
\label{2Dto1D}

We begin with the calculation of the one-dimensional limit for the two-particle
weighted density $N({\bf r})$ defined in Eq.~(\ref{N2}).  
Substituting the expressions $\rho_i({\bf r})=\rho_i(x)\delta(y)$
(where $\rho_i(x)$ is the one-dimensional density of species
$i$) and integrating over the coordinates $y_i$ ($i=1,2$) we obtain
\begin{equation}
\begin{split}
N(x,y)=&
\frac{1}{4\pi^2}\sum_{i,j}\frac{1}{R_iR_j}\int_{-\infty}^{\infty}dx_1
\int_{-\infty}^{\infty}dx_2\,\rho_i(x_1)\rho_j(x_2)
\delta\left(R_i-\sqrt{(x-x_1)^2+y^2}\right) \\
&\times\delta\left(R_j-\sqrt{(x-x_2)^2+y^2}\right)K_{ij}(|x_1-x_2|) \\
=&\frac{1}{4\pi^2}\sum_{i,j}\frac{\Theta(R_i-|y|)\Theta(R_j-|y|)}{u_iu_j}
\int_{-\infty}^{\infty}dx_1\int_{-\infty}^{\infty}dx_2\,\rho_i(x_1)\rho_j(x_2) \\
&\times \delta(u_i-|x-x_1|)\delta(u_j-|x-x_2|)K_{ij}(|x_1-x_2|),
\label{eq:A1}
\end{split}
\end{equation}
where $u_i\equiv\sqrt{R_i^2-y^2}$ and we have used the identity
\begin{equation}
\delta\left(R_i-\sqrt{(x-x_1)^2+y^2}\right)=\Theta(R_i-|y|)\frac{R_i}{u_i}
\delta(u_i-|x-x_1|).
\end{equation}
Because of the deltas in the integral (\ref{eq:A1}), it can be readily performed
and yields
\begin{equation}
\begin{split}
N(x,y)=&\sum_{i,j}\frac{\Theta(R_i-|y|)\Theta(R_j-|y|)}{4\pi^2u_iu_j}
\left\{[\rho_i(x+u_i)\rho_j(x+u_j)+\rho_i(x-u_i)\rho_j(x-u_j)]K_{ij}(|u_i-u_j|)
\right. \\
&\left.
+[\rho_i(x+u_i)\rho_j(x-u_j)+\rho_i(x-u_i)\rho_j(x+u_j)]K_{ij}(u_i+u_j)\right\}.
\end{split}
\end{equation}
From Eqs.~(\ref{kernel}) and (\ref{kernel2}) for $K_{ij}(r)$ we obtain
\begin{equation}
K_{ij}(|u_i\pm u_j|)=\pi|y||u_i\pm u_j||\lambda_i\pm\lambda_j|,
\end{equation}
where we have defined
\begin{equation}
\lambda_i\equiv\sin^{-1}(u_i/R_i)=\cos^{-1}(|y|/R_i).
\end{equation}

In order to proceed let us assume for a while that $R_i\ge R_j$. Then
$u_i\ge u_j$ and $\lambda_i\ge\lambda_j$ and therefore
\begin{equation}
\begin{split}
N(x,y)=&\sum_{i,j}\frac{\Theta(R_i-|y|)\Theta(R_j-|y|)|y|}{4\pi} \\
&\times
\left\{[\rho_i(x+u_i)+\rho_i(x-u_i)][\rho_j(x+u_j)+\rho_j(x-u_j)]
\left(\frac{\lambda_i}{u_j}+\frac{\lambda_j}{u_i}\right) \right. \\
&-\left.
[\rho_i(x+u_i)-\rho_i(x-u_i)][\rho_j(x+u_j)-\rho_j(x-u_j)]
\left(\frac{\lambda_i}{u_i}+\frac{\lambda_j}{u_j}\right) \right\}.
\end{split}
\label{eq:A6}
\end{equation}
In order to get the equivalent expression when $R_i<R_j$ we should just exchange
the indices $i$ and $j$ in the above expression. But this expression is invariant
under this exchange of indices, therefore it holds for any $R_i$ and $R_j$.

Let us now obtain the densities $n_0({\bf r})$ and $n_2({\bf r})$ given by
Eqs.~(\ref{n0}) and (\ref{n2}). When inserting the one-dimensional density
profile one gets
\begin{eqnarray}
n_2(x,y)=\sum_in_{2i}(x,y), &\qquad& n_{2i}(x,y)=\Theta(R_i-|y|)
\int_{x-u_i}^{x+u_i}\rho_i(t)\,dt, \label{eq:A7} \\
n_0(x,y)=\sum_in_{0i}(x,y), 
&\qquad& n_{0i}(x,y)=\frac{\Theta(R_i-|y|)}{2\pi u_i}
[\rho_i(x+u_i)+\rho_i(x-u_i)]. \label{eq:A8}
\end{eqnarray}
Notice that, as $u_i=R_i$ when $y=0$, then $n_2(x,0)=n_1(x)$ and
$n_{0i}(x,0)=n_{0i}(x)/\pi R_i$, with
\begin{equation}
n_1(x)=\sum_i\int_{x-R_i}^{x+R_i}\rho_i(t)\,dt, \qquad
n_0(x)=\sum_in_{0i}(x)
=\frac{1}{2}\sum_i[\rho_i(x+R_i)+\rho_i(x-R_i)], 
\label{eq:1Ddensities}
\end{equation}
the two weighted densities of the exact DF for a mixture of 1D hard rods
\cite{vanderlick:1989}.

For the sake of notational clarity, in what follows we will omit the
arguments of $n_0(x,y)$ and $n_2(x,y)$.
Equations~(\ref{eq:A7}) and (\ref{eq:A8}) help us
to rewrite (\ref{eq:A6}) as
\begin{equation}
\begin{split}
N(x,y)&=\sum_{i,j}\left\{\pi|y|n_{0i}n_{0j}(u_i\lambda_i+u_j\lambda_j)
-\frac{|y|}{4\pi}\frac{\partial n_{2i}}{\partial x}\frac{\partial n_{2j}}{\partial x}
\left(\frac{\lambda_i}{u_i}+\frac{\lambda_j}{u_j}\right) \right\} \\
&=2\pi|y|n_0\sum_iu_i\lambda_in_{0i}-\frac{|y|}{2\pi}
\frac{\partial n_2}{\partial x}\sum_i\frac{\lambda_i}{u_i}
\frac{\partial n_{2i}}{\partial x}
\end{split}
\label{eq:A9}
\end{equation}
(for notational simplicity we have omitted the $x$ and $y$ dependence of the weighted
densities).

Now we can integrate $\Phi^{\rm (2D)}(x,y)$, as given by Eq.~(\ref{fi_2D}), with
respect to $y$ to obtain
\begin{equation}
\begin{split}
\tilde\Phi^{\rm (1D)}(x)&=\int_{-\infty}^{\infty}\Phi^{\rm (2D)}(x,y)\,dy
=\int_{-\infty}^{\infty}dy\,\left\{-n_0(x,y)\ln[1-n_2(x,y)]+
\frac{N(x,y)}{1-n_2(x,y)}\right\} \\
&=2\int_0^{\infty}dy\,\left\{-n_0(x,y)\ln[1-n_2(x,y)]+
\frac{N(x,y)}{1-n_2(x,y)}\right\},
\end{split}
\end{equation}
because the integrand is an even function of $y$. On the other hand, when
$y\ge 0$
\begin{eqnarray}
\frac{\partial n_2}{\partial y} &=& -2\pi yn_0, \\
\frac{\partial}{\partial y}(u_i\lambda_in_{0i}) &=&
-n_{0i}-\frac{\lambda_iy}{2\pi u_i}\frac{\partial^2n_{2i}}{\partial x^2},
\end{eqnarray}
therefore
\begin{equation}
\begin{split}
\tilde\Phi^{\rm (1D)}(x)&=2\int_0^{\infty}dy\,\sum_i\left\{
\frac{\partial}{\partial y}\left[u_i\lambda_in_{0i}\ln(1-n_2)\right]
+\frac{\partial}{\partial x}\left[\frac{\lambda_i y}{2\pi u_i}\ln(1-n_2)
\frac{\partial n_{2i}}{\partial x}\right]\right\}.
\end{split}
\end{equation}
The first term in this equation can readily integrated. Since for
$y=0$ we have $u_i(0)=R_i$ and $\lambda_i(0)=\pi/2$ and for $y=R_i$
we have $u_i(R_i)=\lambda_i(R_i)=0$, it follows that
\begin{equation}
\tilde\Phi^{\rm (1D)}(x)=\Phi^{\rm (1D)}(x)+\frac{\partial\Psi}{\partial x}(x),
\end{equation}
where
\begin{equation}
\Phi^{\rm (1D)}(x)=-n_0(x)\ln[1-n_1(x)]
\label{eq:phi1D}
\end{equation}
is the exact DF for a 1D hard rod mixture \cite{vanderlick:1989} in
terms of the weighted densities (\ref{eq:1Ddensities}), and
\begin{equation}
\Psi(x)=\int_0^{\infty}dy\,\sum_i
\left[\frac{\lambda_i y}{\pi u_i}\ln(1-n_2)
\frac{\partial n_{2i}}{\partial x}\right].
\end{equation}
Assuming proper boundary conditions for the density when $x\to\pm\infty$,
the free-energy functional for the system is given by
\begin{equation}
\beta\mathcal{F}_{\rm ex}^{\rm (2D)}[\{\rho_i\}]=
\int_{-\infty}^{\infty}\tilde\Phi^{\rm (1D)}(x)\,dx=
\int_{-\infty}^{\infty}\Phi^{\rm (1D)}(x).
\label{eq:1Dfunctional}
\end{equation}
This completes the proof of the exact 2D$\to$1D dimensional crossover of the
DF for HD (\ref{fi_2D}).

\section{Geometric measures of the overlap between two cylinders}
\label{c2}

In this appendix we provide explicit expressions for the 
geometrical measures of the body formed by overlapping two 
cylinders with radii $R_i$ and lengths $L_i$. To begin with, the
formal definition of all these measures is
\begin{eqnarray}
f_{ij}({\bf r}_{12}) &=& -\Theta_{R^{(+)}_{ij}}(r_{12}^{\perp})\Theta_{L^{(+)}_{ij}}(z_{12}), \\
\Delta L^{\parallel}_{ij}({\bf r}_{12}) &=& \Theta_{R^{(+)}_{ij}}(r_{12}^{\perp})
\left[\Theta_{L_i}*\Theta_{L_j}(z_{12})\right], \\
\Delta L^{\perp}_{ij}({\bf r}_{12}) &=& \left\langle\Theta_{R_i}*\delta_{R_j}
(r_{12}^{\perp})\right\rangle\Theta_{L^{(+)}_{ij}}(z_{12}), \\
\Delta S^{\parallel}_{ij}({\bf r}_{12}) &=& 2\left[\Theta_{R_i}*\Theta_{R_j}
(r_{12}^{\perp})\right]\Theta_{L^{(+)}_{ij}}(z_{12}),
\end{eqnarray}
where $\Theta_u(r^{\perp})=\Theta(u-r^{\perp})$,
$\Theta_u(z)=\Theta(u/2-z)$, and $\delta_u(r^{\perp})=\delta(u-r^{\perp})$.
It is rather easy to evaluate these expressions appealing to their
geometrical meaning. Hence, the total arch lengths of 
the cross-section of the overlap body is 
\begin{eqnarray}
\Delta L_{ij}^{\perp}(r_{12}^{\perp})=2\left\{
\left\langle R_i\cos^{-1}t_{ij}\right\rangle\Theta\left(r_{12}^{\perp}
-R_{ij}^{(-)}\right)+
\frac{\pi}{2}\left(R_{ij}^{(+)}-R_{ij}^{(-)}\right)
\Theta\left(R_{ij}^{(-)}-r_{12}^{\perp}\right)\right\},
\end{eqnarray}
while its height is
\begin{eqnarray}
\Delta L_{ij}^{\parallel}(z_{12})=\frac{L_{ij}^{(+)}}{2}-|z_{12}|-
\left(\frac{L_{ij}^{(-)}}{2}-|z_{12}|\right)\Theta\left(L_{ij}^{(-)}/2-|z_{12}|\right).
\end{eqnarray}
Similarly, the expression for twice the area of the base of the overlap body is
given by
\begin{eqnarray}
\Delta S_{ij}^{\parallel}(r_{12}^{\perp})&=&2\Bigg\{\left\langle R_i^2
\left[\cos^{-1} t_{ij}-\frac{r_{12}^{\perp}}{2R_i}
\sqrt{1-t_{ij}^2}\right]\right\rangle
\Theta\left(r_{12}^{\perp}-R_{ij}^{(-)}\right)\nonumber\\
&&+\frac{\pi}{4}\left(R_{ij}^{(+)}-R_{ij}^{(-)}\right)^2
\Theta\left(R_{ij}^{(-)}-r_{12}^{\perp}\right)\Bigg\},
\end{eqnarray}
while its lateral area is readily obtained as
\begin{eqnarray}
\Delta S_{ij}^{\perp}({\bf r}_{12})=\Delta L_{ij}^{\perp}(r_{12}^{\perp})
\Delta L_{ij}^{\parallel}(z_{12}).
\label{b4}
\end{eqnarray}

%\bibliography{liquid}

%\end{document}

\end{document}